\documentclass[pre,notitlepage,showpacs,amssymb,amsfonts,amsmath,floatfix]{revtex4-1}

\usepackage{natbib}
\usepackage{hyperref}
\usepackage{graphicx}
\usepackage{epsfig}
\usepackage{dcolumn}
\usepackage{bm}
\usepackage{times}

\begin{document}
\title{Multi-surface coding simulations of the restricted solid-on-solid
model in four dimensions}
\author{Andrea Pagnani}
\affiliation{Human Genetics Foundation (HuGeF), Via Nizza 52, I-10126, Turin, Italy}
\author{Giorgio Parisi}
\affiliation{Dipartimento di Fisica, INFN - Sezione di Roma 1, CNR-IPCF UOS Roma, Universit\`a ``La Sapienza'', P.le Aldo Moro 2, I-00185 Roma, Italy}


\begin{abstract}
We study the Restricted Solid on Solid (RSOS) model for surface growth
in spatial dimension $d=4$ by means of a {\em multi-surface coding}
technique that allows to analyze samples of size up to $256^4$ in the
steady state regime. For such large systems we are able to achieve a
controlled asymptotic regime where the typical scale of the
fluctuations are larger than the lattice spacing used in the
simulations. A careful finite-size scaling analysis of the critical
exponents clearly indicate that $d=4$ is not the upper critical
dimension of the model.
\end{abstract}

\maketitle 

The Kardar-Parisi-Zhang (KPZ) equation \cite{KPZ86} is possibly the
simplest, most studied, and yet not fully understood model for
out-of-equilibrium surface growth. The equation describes the time
evolution of the height $h({\mathbf r}, t)$ of an interface above a
$d-$dimensional substrate:
\begin{equation}
\label{eq:KPZ}
\partial_t h({\mathbf r}, t) = \nu {\vec \nabla}^2 h({\mathbf r}, t) +
\frac \lambda 2 | {\vec \nabla}h({\mathbf r}, t)|^2 +\eta({\mathbf r}, t) \, ,
\end{equation}
where $\nu$ is the diffusion coefficient, $\lambda$ is the strength of
the non-linear growth rate term which is responsible for the $h
\rightarrow -h$ symmetry breaking with respect to the growing
direction, and $\eta({\mathbf r},t)$ is a Gaussian white noise of
amplitude $D$:
\begin{equation}
\label{eq:noise}
\langle \eta \rangle = 0 \,\,\,\,\,\,,\,\,\,\,\,\,
\langle \eta({\mathbf r},t) \eta({\mathbf r}',t') \rangle 
= 2D \delta^d({\mathbf r} - {\mathbf r}')
\delta(t-t')\,\,. 
\end{equation}

The KPZ equation describe many relevant growth processes, such as the
Eden model, ballistic deposition, interface growth in disordered
medium. It is also related to many other physical phenomena apparently
unrelated to surface growth, such as Burgers turbulence, dynamics
directed polymers in random media, dissipative transport in the
driven-diffusion equation \cite{FamilyVicsekBook, *BarabasiBook}.

The scaling properties of the height's fluctuations $w_2(L,t) =
\langle h^2 ({\mathbf r}, t) \rangle_{\mathbf r} - \langle h ({\mathbf
  r}, t) \rangle^2_{\mathbf r}$ (with the notation $\langle \dots
\rangle_{\mathbf r}$ we indicate a spatial average over a macroscopic
hypercubic box of linear size $L$ over the $d-$dimensional substrate)
characterize the universality class of the model. More precisely, for
a system of size $L$, $w_2(L,t) \sim L^{2 \chi} f(t/L^z)$, where the
scaling function is such that $f(x) \rightarrow \mathrm{const}$ for $x
\rightarrow \infty$ and $f(x) \sim x^{2\chi/z}$ for $x \rightarrow
0$. The peculiar behavior of $f$ imply that $w_2(L,t) \sim L^{2\chi}$
for $t \gg L^z$ and $w_2(L,t)\sim t^{2 \chi/z}$ for $t \ll L^z$. Due
to an infinitesimal tilt symmetry of Eq.~(\ref{eq:KPZ}) ($h\rightarrow
h+{\mathbf r }\cdot{\boldsymbol \epsilon}$, ${\mathbf r} \rightarrow
{\mathbf r} - \lambda t {\boldsymbol \epsilon} $), the two critical
exponents are related by the scaling relation $\chi + z = 2$, which is
believed to be valid at any dimension $d$ \cite{FamilyVicsekBook,
  *BarabasiBook}.

A complete understanding of Eq.~(\ref{eq:KPZ}), and in particular the
determination of the two critical exponents $\chi,z$ for any spatial
dimension $d>1$ (at $d=1$ a fluctuation-dissipation theorem leads to
the exact result $\chi = 1/2$, $z=3/2$) , turns out to be extremely
difficult for two main reasons: (i) we are dealing with an
intrinsically out-of-equilibrium phenomenon where the standard
equilibrium toolbox must be used with care, (ii) perturbative
renormalization schemes are not adequate for describing the strong
coupling regime ({\em i.e.}~where the parameter $\lambda$ is
relevant). 

The existence of an upper critical dimension $d_u$, {\em i.e.} the
substrate dimensionality $d$ above which the fluctuation of the model
become irrelevant ($\chi = 0$), is one of the most controversial
unsolved theoretical issues related with Eq.~(\ref{eq:KPZ}). The
determination of $d_u$ would be a most relevant achievement since, as
customary in equilibrium critical phenomena, its knowledge constitutes
the first step for a controlled perturbative expansion around it. The
quest for $d_u$ has been around for more than twenty years
\cite{Halpin-Healy1990,Cook-Derrida1990,%
  Schwartz-Edwards1998,Bouchaud-Cates1993,Bouchaud-Cates-Errata1993,%
  DMKB1994,MBDMBC95, Tu1994, ColaioriMoore2001, Laessig1995,%
  Laessig-Kinzelbach1997,Laessig-Kinzelbach1998, Katzav2002,
  NOIKPZ2001, *NOIKPZ2002, Frogedby2006, Fogedby2008,%
  CCDW2010, KK1989, *WK1987,
  *TLFWD1992, *Nissila1993, CMP1998,%
  *CGMMP1998,*CMMP1999} and the
  different predictions range from $d_u \approx 2.8$ to $d_u =
  \infty$. Analytical estimates using the mode-coupling theory yield
  exact results in $d=1$ \cite{HwaFrey1991,*FTUH1996}. Their extension
  to higher dimensions hints for a $d_u = 4$ under different
  self-consistency schemes
  \cite{Schwartz-Edwards1998,Bouchaud-Cates1993,*Bouchaud-Cates-Errata1993,
    DMKB1994,*MBDMBC95,Tu1994, ColaioriMoore2001}. The same value for
  $d_u$ is also supported by different field-theoretic approaches
  \cite{Halpin-Healy1990, Laessig1995,*Laessig-Kinzelbach1997,Laessig-Kinzelbach1998, Frogedby2006, Fogedby2008}, whereas in \cite{CCDW2010}
  a non-perturbative renormalization group technique is proposed
  yielding a finite (although very small compared with numerical
  simulations) scaling coefficient $\chi$ in $d=4$.

At odd with what predicted by the previously mentioned field-theoretic
approaches, both direct numerical integration of KPZ equation
\cite{Moser1991}, and simulation of systems belonging to the KPZ
universality class \cite{
  KK1989,WK1987,TLFWD1992,Nissila1993,Ala-Nissila98,NOIKPZ2001,*NOIKPZ2002,
  Perlsman-Havlin2006,OrdorLiedkeHeinig-PRE2010,Schwartz-Perlsman2011,KellingOrdor2011}
indicate that $d_u > 4$, while the real-space renormalization group
approach \cite{ CMP1998, *CGMMP1998,*CMMP1999} predicts $d_u =
\infty$.

Such a long standing controversy is the consequence of the
difficulties inherent to both analytical and numerical
approaches. Most of the assumptions made on the functional structure
of the sought solution in the different field-theoretic analysis, as
well as the approximations made in the mode-coupling theories are, in
general, not completely under control. On the numerical side the most
severe problem is due to the fact that simulations in high spatial
dimensions $d \geq 4$ are computationally very heavy, and the systems
under analysis must be limited in size. As a consequence, the
different fitting procedures must deal with controlled finite-size
scaling procedures to yield reliable estimates of the critical
exponents. Under this perspective, particularly relevant is the
observation that for lattice models in the KPZ universality class, a
controlled asymptotic regime is achieved only when typical scale of
the fluctuations is larger than the lattice spacing used in the
simulations or, more precisely, for $w_2 > 1$
\cite{ColaioriMoore2001}. The former inequality is very stringent from
the computational point of view since it requires very large lattices
to be fulfilled: the estimates presented in \cite{NOIKPZ2001,
  OrdorLiedkeHeinig-PRE2010} suggest indeed that for the
$4-$dimensional RSOS model, to which this letter is addressed, the
$w_2>1$ inequality starts being verified for lattice size larger then
$L \approx 32$, whereas the larger system size analyzed in the steady
state regime is $L=128$ \cite{OrdorLiedkeHeinig-PRE2010} that, at the
best of our knowledge, remains the larger system in 4 dimensions
simulated so far.

To settle the controversy, at least in the $4-$dimensional RSOS model
case, we decided to analyze samples of unprecedented size: we have
been able investigate the steady state scaling regime $t\gg L^z$ for
lattice size volumes up to $V=128^4=268435456$ sites, a factor 16 off
the largest simulation known in literature
\cite{OrdorLiedkeHeinig-PRE2010} in the steady state regime. We can
also study the dynamic scaling regime of lattices of $V = 256^4 =
4294967296$ sites, but for such a large size we have been able to
investigate the asymptotic scaling regime for just three samples, due
to limitations in our computing facility (most of the data for $L=256
$ are at not too large $t$ and they have been used only in fig. (1,2):
they appear only in the region $t/L^{z}<8$).

The RSOS can be simulated in the following way: at any time $t$ we
randomly select a site $i$ on the $d-$dimensional lattice and we let
the surface height $h_i$ at that point to grow of a unit $h_i(t+1) =
h_i(t)+1$ only if $\max_{j \in \partial i} | h_i(t) - h_j(t)| \leq 1$,
being $\partial_i$ the set of 8 nearest neighbors of $i$ in $d=4$
(note that we will assume periodic boundary conditions).

We simulated RSOS growth using two different algorithms based on a
very efficient {\em multi-spin coding} technique \cite{Rieger93}
originally developed for disordered spin system, and later generalized
to deal with the RSOS model \cite{NOIKPZ2001}:
\begin{itemize}
\item {\em Multi Surface (MS) coding:} we can simulate, with basically
  the same cost of one single surface simulation, $N_{b}$ copies of
  the system, $N_{b}$ being the number of bits in an computer world
  (usually 32, 64, 128 and 256). We transform the basic operations
  (like summing spins for computing the effective force) into Boolean
  operations, and we exploited the fact that when, for instance, the
  computer is calculating an AND logical bit, it is indeed doing that
  operation $N_{b}$ times at once, i.e. for all the bits of the
  world. Unfortunately computational efficacy of this method is
  counterbalanced by the memory load, making it unpractical for
  analyzing samples of linear size larger than $L = 64$, at least on
  the computers we have access to.

\item {\em Multi Lattice-site (ML) coding}: for sample of linear size
  $L=128,256$ we have developed a new multi-spin coding representation
  in which a single surface at time is simulated, but we lump together
  $N_{b}$ height-difference local variable in a single computer word
  of $N_{b}$ bits. We will refer to this second method as the Multiple
  Lattice-site (ML) algorithm (see table~\ref{tab:simulpara}).  In
  this algorithm in the first half step we update the even spins
  (i.e. the spins $\sigma_{i}$ where $i_{x}+i_{y}+i_{z}+i_{t}$ is
  even) applying the standard RSOS procedure with probability 1/2 to
  each spin (with probability 1/2 the spin is not updated), in the
  second half step we apply the algorithm to odd spins. Some
  programming care must be used with periodic boundary conditions: in
  the simplest version we have used, $L$ must be an even multiple of
  $N_{b}$. Moreover it is crucial to use a good random number
  generator, where {\it all} the bits are random.
\end{itemize}

\begin{center}
\begin{table}[htb]
\begin{tabular}{|*{5}{c|}}
\hline
 $L$ & sweeps  & samples & type & time (h.)\\
\hline
\hline
8 & 524000 & 1024 & MS & 4 \\
\hline
16 & 524000 & 1024 & MS & 6 \\
\hline
31 & 524000 &  1024 & MS &121 \\
\hline
32 & 524000 & 1024 & MS & 139 \\
\hline
33 & 524000 & 1024 & MS & 158 \\
\hline
64 & 131000 & 512 & MS  & 5376 \\
\hline
128 & 512000 & 64 & ML & 7680 \\
\hline
256 & 130000 & 64 & ML & 504 \\
\hline
\end{tabular}
\protect\caption{In this table we display the lattice linear size $L$,
  the number of montecarlo sweeps (full lattice update), the number of
  samples and the simulation type (MS = multi-surface coding, ML =
  multi-lattice coding), and the overall computational time in hours.}
\label{tab:simulpara}
\end{table}
\end{center}

We simulate $4-$dimensional lattices of volume $V = L^4$ for lattices
of linear size $L=8,16,31,32,33,64,128,256$. For the two largest lattice
($L=64,128,256$) we run the ML algorithm, while for the rest we run the MS
algorithm. We decided to consider $L=31,33$ for checking that there are no
periodicity issues with the random number generator.  A summary of our
simulations is provided in table~\ref{tab:simulpara}.

\begin{widetext}
\begin{center}
\begin{table}[htb]
\begin{tabular}{|*{9}{l|}}
\hline
  & $\chi$ & $\omega$ & $A_2$ & $B_2$  & $A_3$ &  $B_3$ & $A_4$ & $B_4$ \\
\hline
\hline
NEW & 0.2537(8)  & 1.11(9) & 0.171(1)& 0.37(6)& 0.0319(3)& -1.0(2) & 0.100(1)&  0.38(8) 
\\
\hline
OLD &  0.255(3) &0.98(9) & 0.170(1)& 0.37(3) & 0.0321(2) & -0.7(1) & 0.100(1) & 0.46(4) 
\\
\hline
\end{tabular}
\protect\caption{In this table we display the best fit values together
with their statistical error of the parameters defined in
Eq.~(\ref{eq:fit}). The first row refers to the actual data presented
in this work, the second is taken from \cite{NOIKPZ2001}. The value
for $\chi$ is in good agreement with the result $\chi=0.245(5)$
reported in \cite{OrdorLiedkeHeinig-PRE2010} for the directed $4-$mer 
diffusion model.}
\label{tab:fitpara}
\end{table}
\end{center}
\end{widetext}

The simulations aim at achieving a fair sampling of the asymptotic
regime. To do so, at any time $t$ and for each sample (both in ML and
MS type of simulation) we evaluate the first three connected moments
$w_n(L,t) = \sum_{i=1}^V (h_i(t) - \langle h(t)\rangle)^n/V$, where
$\langle h(t) \rangle = \sum_{i=1}^V h_i(t)/V$, and $n=2,3,4$. We thus
define our asymptotic (in time) estimate as:
\begin{equation}
w_n(L) = \frac1{T_0-T_1+1} \sum_{t=T_1}^{T_0}w_n(L,t )\,\,\,\,.
\end{equation}

\begin{figure}[ht]
\begin{center}
\includegraphics[width=\columnwidth]{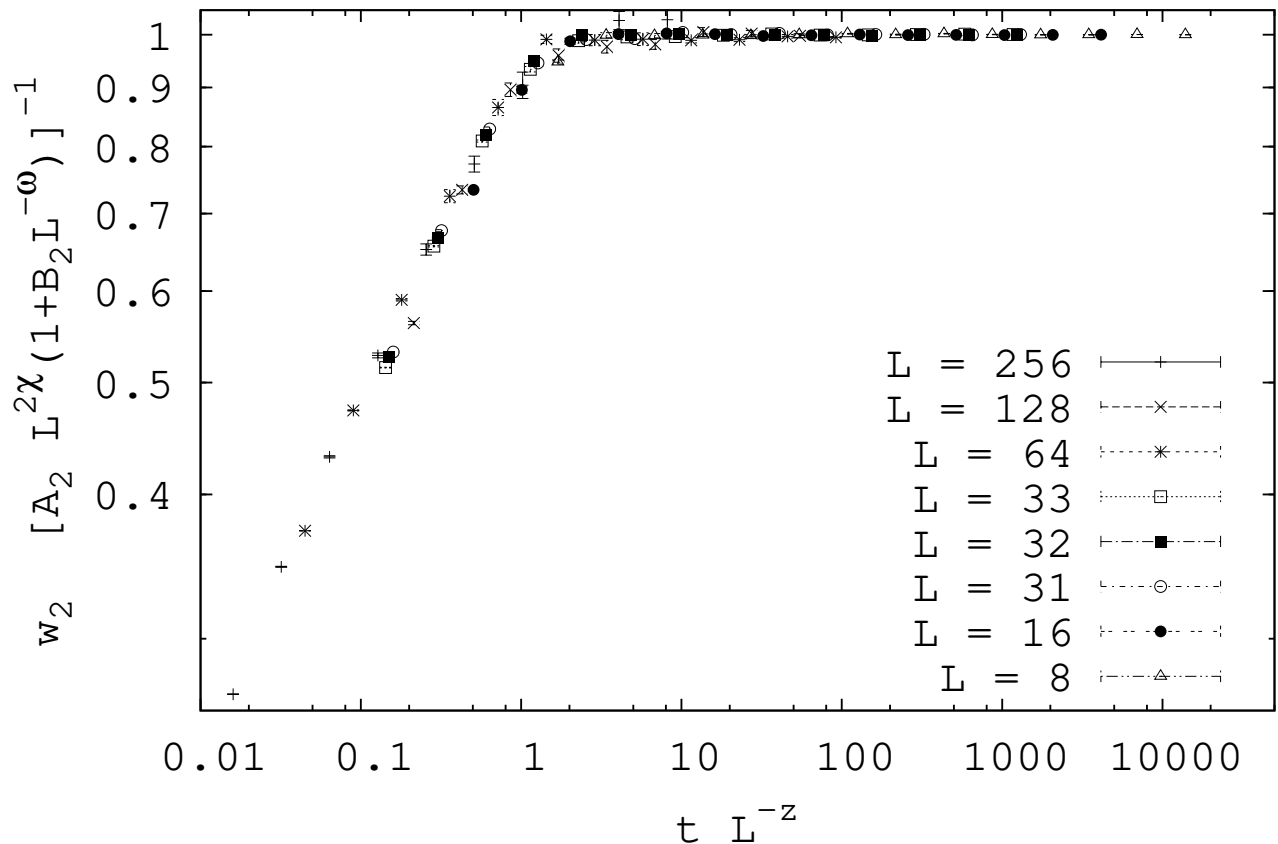}
\end{center}
\caption{\label{fig:scaling} Scaling plot of the rescaled second
  moment $w_2/(A_2L^{2\chi}(1+B_2L^{-\omega}))$ vs. the rescaled time $t/L^z$.}
\end{figure}

We are careful to choose both $T_0$, and $T_1-T_0$ large enough to
guarantee: (i) convergence to the asymptotic regime, {\em~i.e.} that
$T_1 \gg L^z$, (ii) a large enough sampling of statistically
uncorrelated measures of $w_n(L,t)$. In practice we consider
consecutive measuring windows of length $64, 128, ...,T_1,T_0$, so
that $T_0$ is the last measure in the simulation and $T_1 =
T_0/2$. This choice is very conservative, since eventually we use just
the second half of the simulation, but at the same time it allows us
to check with very high reliability whether or not we have reached the
asymptotic state: a quick look at Fig.~\ref{fig:scaling} will comfort
our confidence of having chosen for $L=128$ a $T_0 \gg L^z$, at least
of a factor 100 off, whereas for $L=256$, due to the computational
cost, only a factor $10$ off. We are now ready to determine the
critical exponents of the asymptotic behavior of $w_n$, which scales
as $L^{n\chi}$ at the leading order, by fitting simultaneously the
following moments \cite{ChinDenNijs99,NOIKPZ2001}:
\begin{eqnarray}
\label{eq:fit}
w_2  &=& A_2 L^{2\chi}( 1 + B_2 L^{-\omega})\nonumber\\
w_3  &=& A_3 L^{3\chi}( 1 + B_3 L^{-\omega})\\
w_4  &=& A_4 L^{4\chi}( 1 + B_4 L^{-\omega})\nonumber\,\,\,\,\,, 
\end{eqnarray}

\begin{figure}[ht]
\begin{center}
\includegraphics[width=\columnwidth]{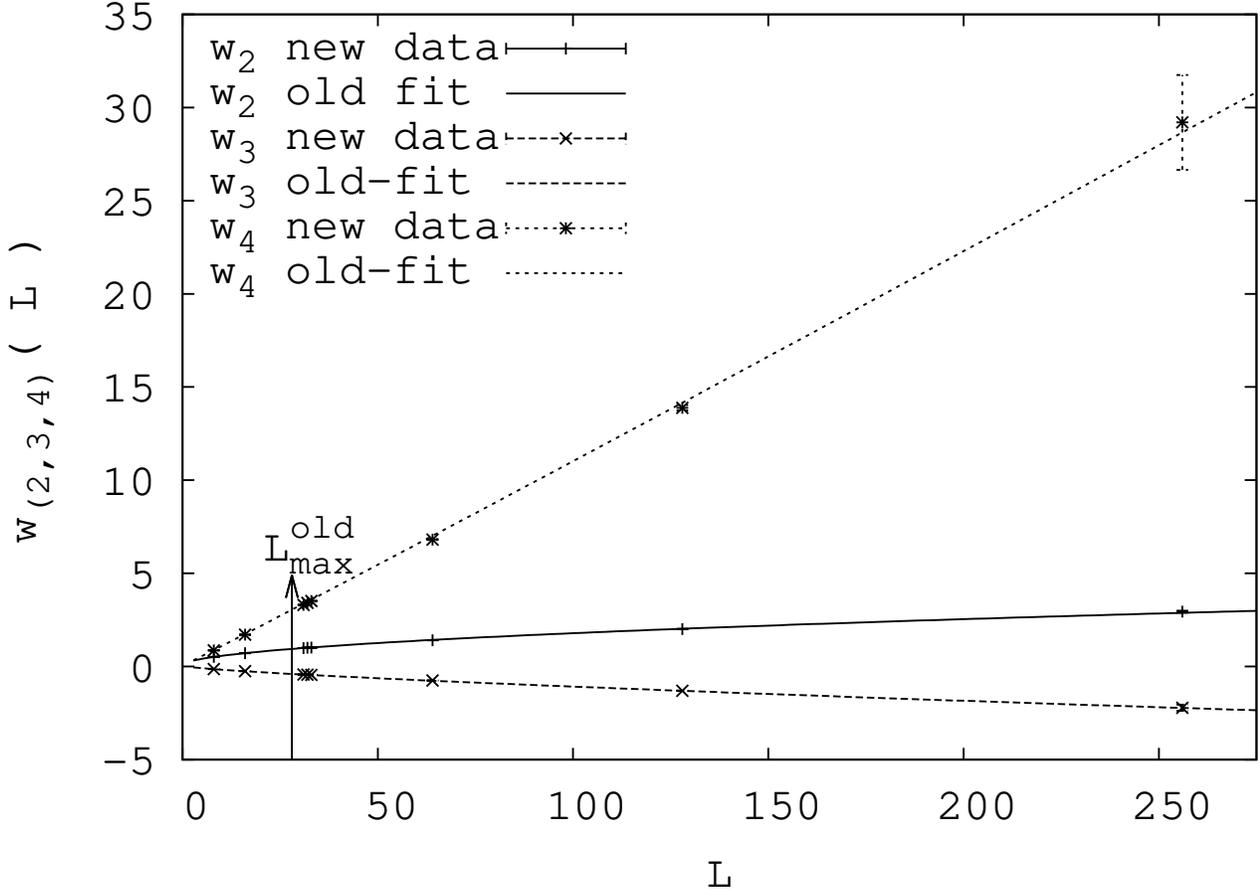}
\end{center}
\caption{\label{fig:fit} The quantities $w_{2},\ w_{3}\ w_{4}$ are
  displayed as a function of $L$. Dots with error bars are values
  obtained by simulations, while solid lines are the 8-parameters
  best-fit reported in \cite{NOIKPZ2001}. The vertical arrow arrow at
  $L=28$ represents the largest size simulated in \cite{NOIKPZ2001}. }
\end{figure}

where $\omega$ is the leading finite size scaling exponent. The fit
involves the simultaneous determination of 8 parameters whose best-fit
value is reported in the first row of table~\ref{tab:fitpara} (the fit
yields a chi-squared root mean square deviation of
2.05). Interestingly enough the fit presented in \cite{NOIKPZ2001}
agrees very well with the new data as we can clearly appreciate
qualitatively in Fig.~\ref{fig:fit}, and more quantitatively by
comparing the two rows in table~\ref{tab:fitpara}. With respect to the
$w_2>1$ inequality issue, a glance at Fig.~\ref{fig:fit} shows
unambiguously that the scale of the typical fluctuations, for all
lattice size larger than $L=31$, verify the inequality. We do not see
any change in the scaling behavior of the three cumulants around the
cross-over region $L\approx 30$, moreover, the fact that the old fit
presented in \cite{NOIKPZ2001} (in a regime $w_2<1$) agrees so well
with our larger lattice size simulation (see again Fig.~\ref{fig:fit})
indicate that the simulations performed for $L\leq 28$ were able to
capture fairly the asymptotic scaling regime. To see more clearly the
finite-size corrections of $\chi$ we determined the effective exponent
$\chi_n^\mathrm{eff}$ as the discretized logarithmic derivative of
Eqs.~(\ref{eq:fit}) which in our case reads:
\begin{equation}
\label{eq:chieff}
\chi_n^\mathrm{eff} (L) = \frac{\log(\frac{w_n(L)}{w_n(L')})}{n \log(\frac L{L'})}
\end{equation}

\begin{figure}[ht]
\begin{center}
\includegraphics[width=\columnwidth]{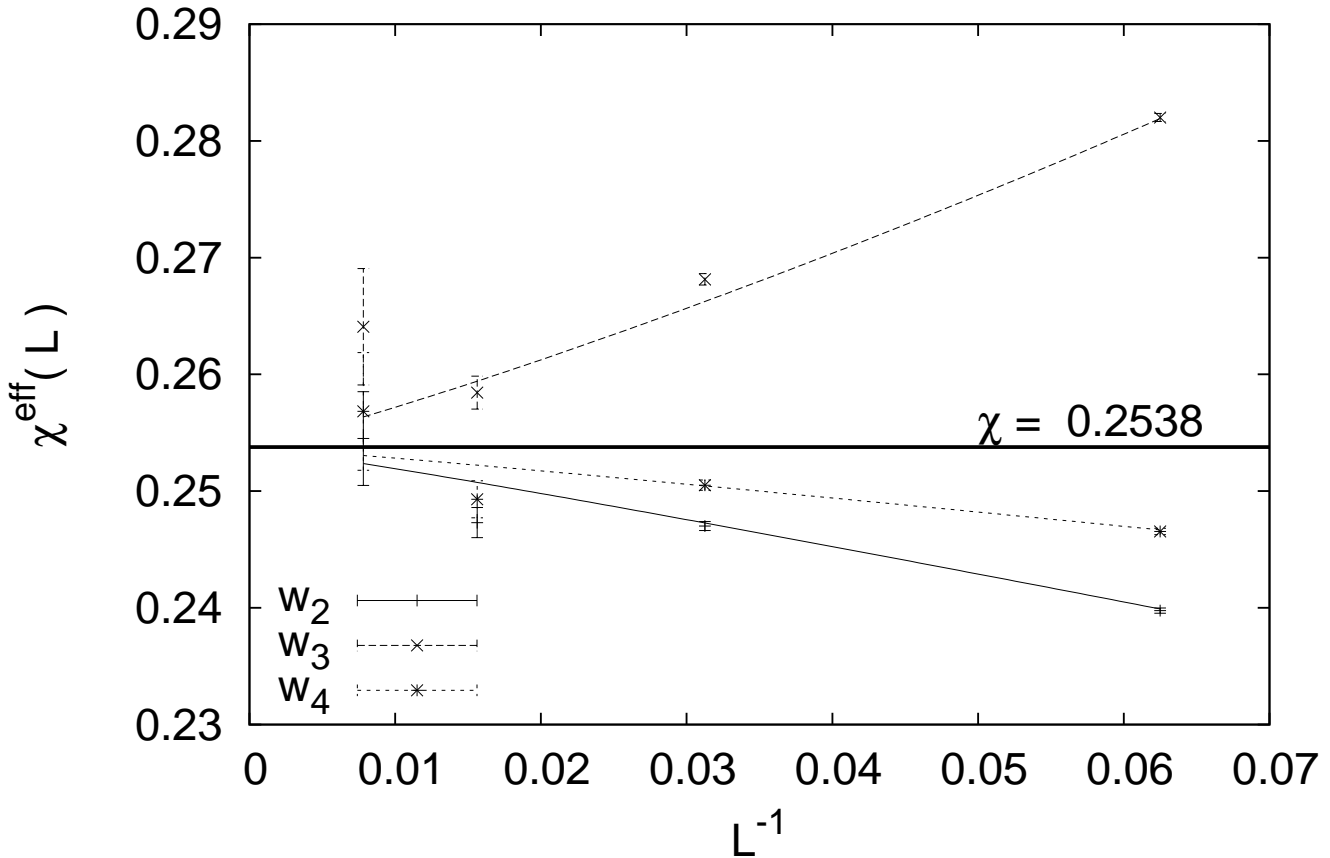}
\end{center}
\caption{\label{fig:localchi} Local slopes of $w_{2,3,4}$ are
  displayed as a function of $L^{-1}$. Dots with error bars are values
  obtained by simulations, while lines are the 8-parameters best-fit
  reported in table~\ref{tab:fitpara}. The solid horizontal line is at
  $\chi=0.2538$, i.e. the best-fit prediction for the wandering
  exponent.}
\end{figure}
where $L/L'= 2$, and $n=1,2,3$. In Fig.~\ref{fig:localchi} we display
$\chi_n^\mathrm{eff}$ as a function of $L^{-1}$ (note that we
discarded for the sake of clarity the $L=31,33$ results) for the three
cumulants together with the best-fit curves. The fit yields the
following results for the critical exponents: $\chi=0.2532(5)$ and
$\omega = 1.14(5)$ (see also table~\ref{tab:fitpara}). Our estimate
for the $\chi$ exponent compare very well with the recent result
obtained in \cite{OrdorLiedkeHeinig-PRE2010} for the directed $4-$mer
diffusion model where a value $\chi=0.245(5)$ is reported. A recent
work has investigated a model of direct polymers in random medium that
should belong to the same universality class
\cite{Schwartz-Perlsman2011}. In this model one can define an exponent
$\zeta$ that according to the theoretical expectations should be given
 
 \begin{equation}
\zeta={1\over 2-\chi}
\end{equation}
Their results ($\zeta$ slightly larger than 0.57), is well consistent
with our prediction $\zeta=0.5725(2)$.

The numerical technique we have introduced has allowed us to run very
precise numerical simulations of the RSOS model in $d=4$ on
unprecedented system size with a limited amount of computational
time. Thanks to the accuracy of the simultaneous measurement of the
three cumulants, the claim that $d=4$ is the upper critical dimension
for systems in the KPZ universality class has to be
rejected. Moreover, the typical fluctuation's length-scale of our
simulations on samples of linear size $L=128$ and $L=256$ are larger
than the lattice spacing, and this is a clear indication that: (i) the
system reached a controlled scaling regime, (ii) the measured scaling
exponents are reliable and not affected by a pre-asymptotic cross-over
regime.

There is still a remote possibility that our data are consistent with
an upper critical dimension $d_u = 4$ of the KPZ equation if we drop
the hypothesis that RSOS in $d=4$ belongs to the KPZ universality
class. However, apart from some work in the past
\cite{Katzav_Schwartz2004}, this hypothesis does not seem to have
support in the mainstream literature on KPZ.
\\ 
\\ 
We thank Enzo Marinari, and Massimo Bernaschi for interesting
discussions, Moshe Schwartz for relevant correspondence, and Marco
Zamparo for reading the manuscript. The numerical simulations
presented here were run using facility of the Human Genetics
Foundation. The European Research Council has provided financial
support through ERC grant agreement no. 247328.

\end{document}